\documentclass{ws-procs9x6-cpt16}
\begin{document}

\newcommand{\refeq}[1]{(\ref{#1})}
\def\etal {{\it et al.}}

\title{Status Report: A Detector for Measuring the\\
Ground State Hyperfine Splitting of Antihydrogen}

\author{B.\ Kolbinger}

\address{Stefan Meyer Institute for Subatomic Physics,
Vienna 1090, Austria}

\author{On behalf of the ASACUSA Collaboration}

\begin{abstract}
The ASACUSA (Atomic Spectroscopy And Collisions Using Slow Antiprotons) collaboration at the Antiproton Decelerator at CERN aims to measure the ground state hyperfine structure of antihydrogen. A Rabi-like spectrometer line has been built for this purpose.
A detector for counting antihydrogen is located at the end of the beam line.
This contribution will focus on the tracking detector, 
whose challenging task it is to discriminate between background events and antiproton annihilations originating from antihydrogen atoms which are produced only in small amounts.
\end{abstract}

\bodymatter

\section{Introduction}
To date no violation of the symmetry of CPT 
has been observed.
The CPT theorem predicts that particles and antiparticles have identical (or sign-opposite) properties
and therefore, in case of atoms, the same characteristic spectrum.
Nevertheless the matter and antimatter asymmetry in the Universe is quantitatively unexplained.
The ASACUSA collaboration aims to investigate this issue by measuring the ground state hyperfine transitions in antihydrogen, the simplest antiatom. 
The first ingredients of antihydrogen atoms, the antiprotons, are extracted from the CERN Antiproton Decelerator (AD) facility, and are accumulated in a Penning trap. The trapped antiprotons are then combined with positrons inside a so-called CUSP trap
and antihydrogen will be formed in a ''mixing'' process.\cite{cusp1}
The neutral antihydrogen atoms escape the trapping field and enter the spectroscopy apparatus.
The Rabi-like spectrometer line\cite{Widmann1} consists of a microwave cavity for inducing spin flips, a superconducting sextupole magnet for analyzing the spin state and a detector for counting antihydrogen atoms composed of a central detector, where the annihilation takes place, and a two layer hodoscope for tracking the charged annihilation products. The central detector is a 10~cm diameter bismuth germanate (BGO) disc
with a thickness of 5~mm. The BGO crystal is read by multi anode photomultipliers  providing energy and position information of the events.\cite{nagatabgo}

\section{Antihydrogen tracking detector --- the hodoscope}  
\begin{figure}[b]
\centering
$\begin{array}{rl}
    \includegraphics[width=0.45\hsize]{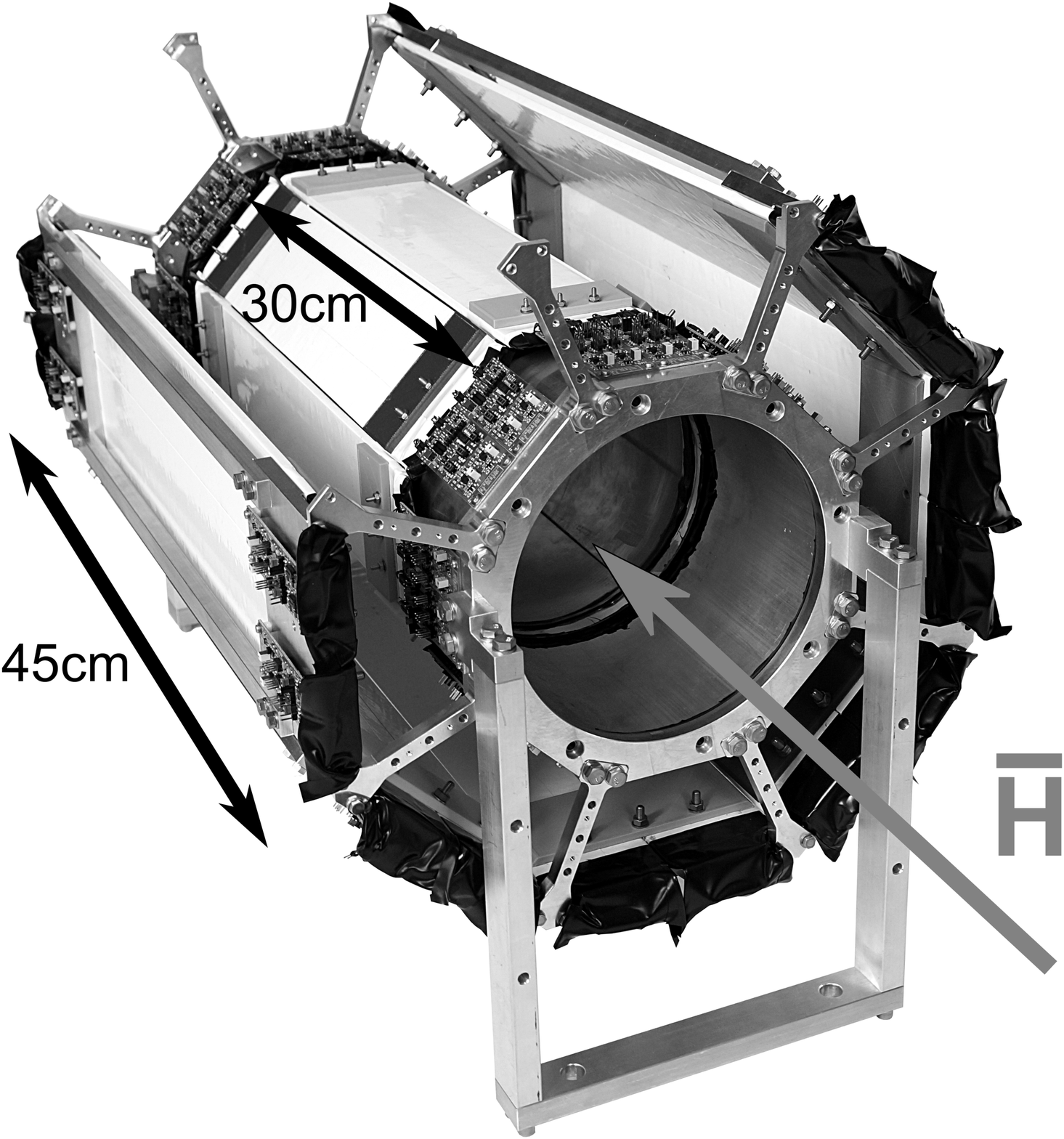} 
    \includegraphics[width=0.6\hsize]{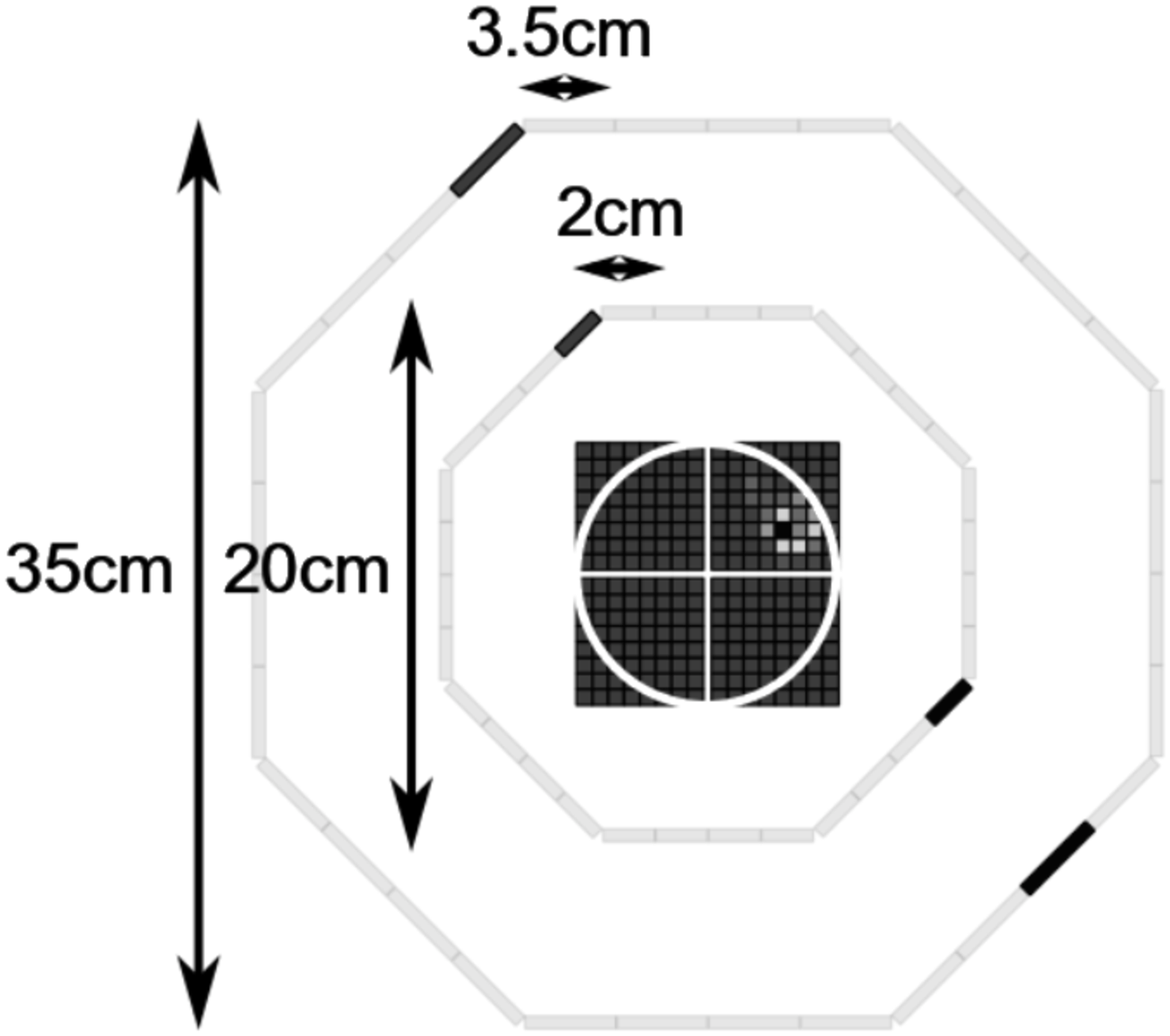}
\end{array}$
\caption{\label{fig:hodor}Left: the two layered hodoscope. Two panels of the outer layer are omitted to show the internal layer. Right: Cross section of the detector perpendicular to the beam showing a cosmic event. The squared pixel map shows the data recorded by the MAPMTs. The BGO is indicated by a white circle. The octogons show the scintillator geometry.}
\end{figure}

\begin{figure}[t]
\centering
$\begin{array}{rcl}
    \includegraphics[width=3.45cm]{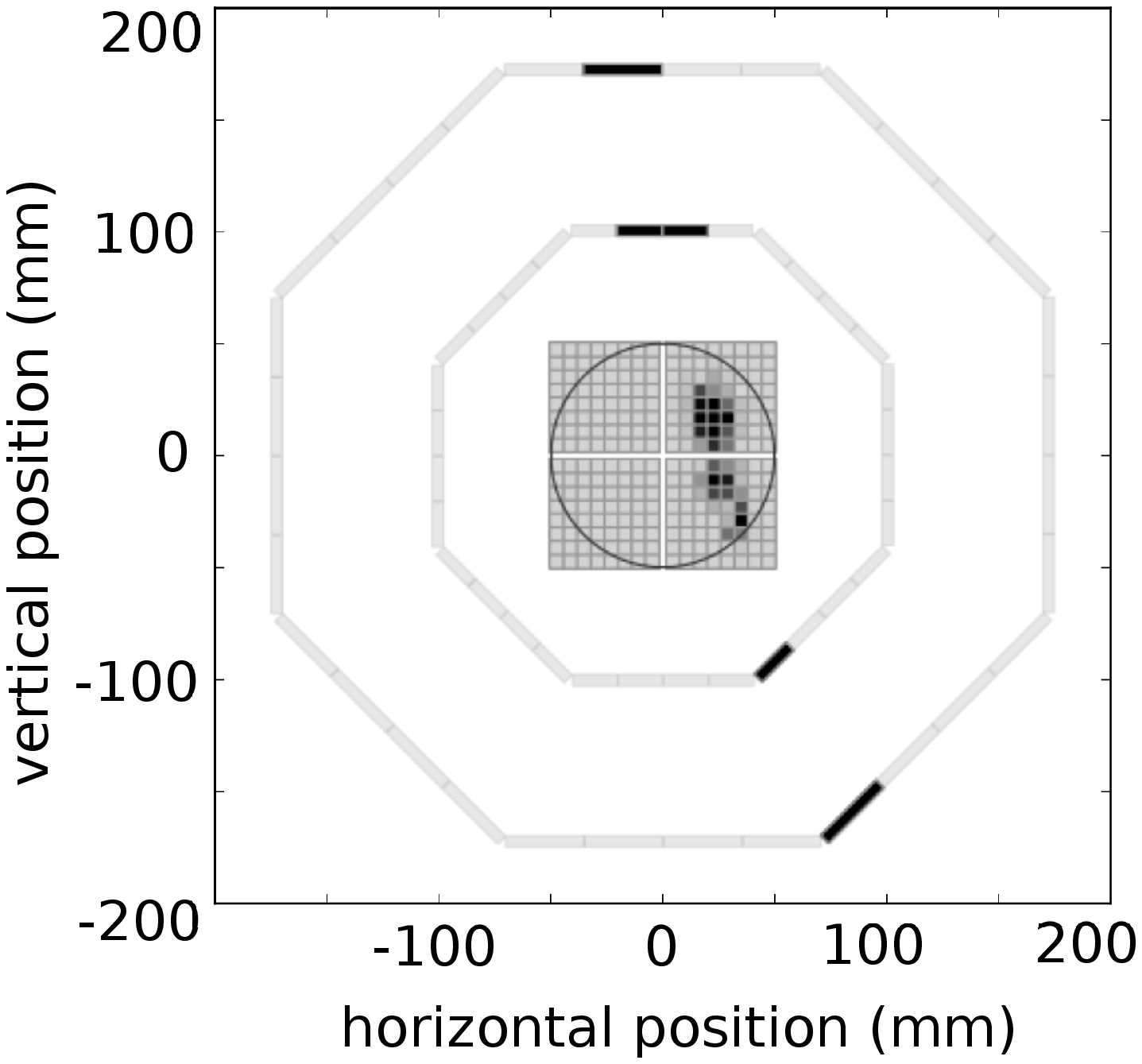} 
    \includegraphics[width=3.5cm]{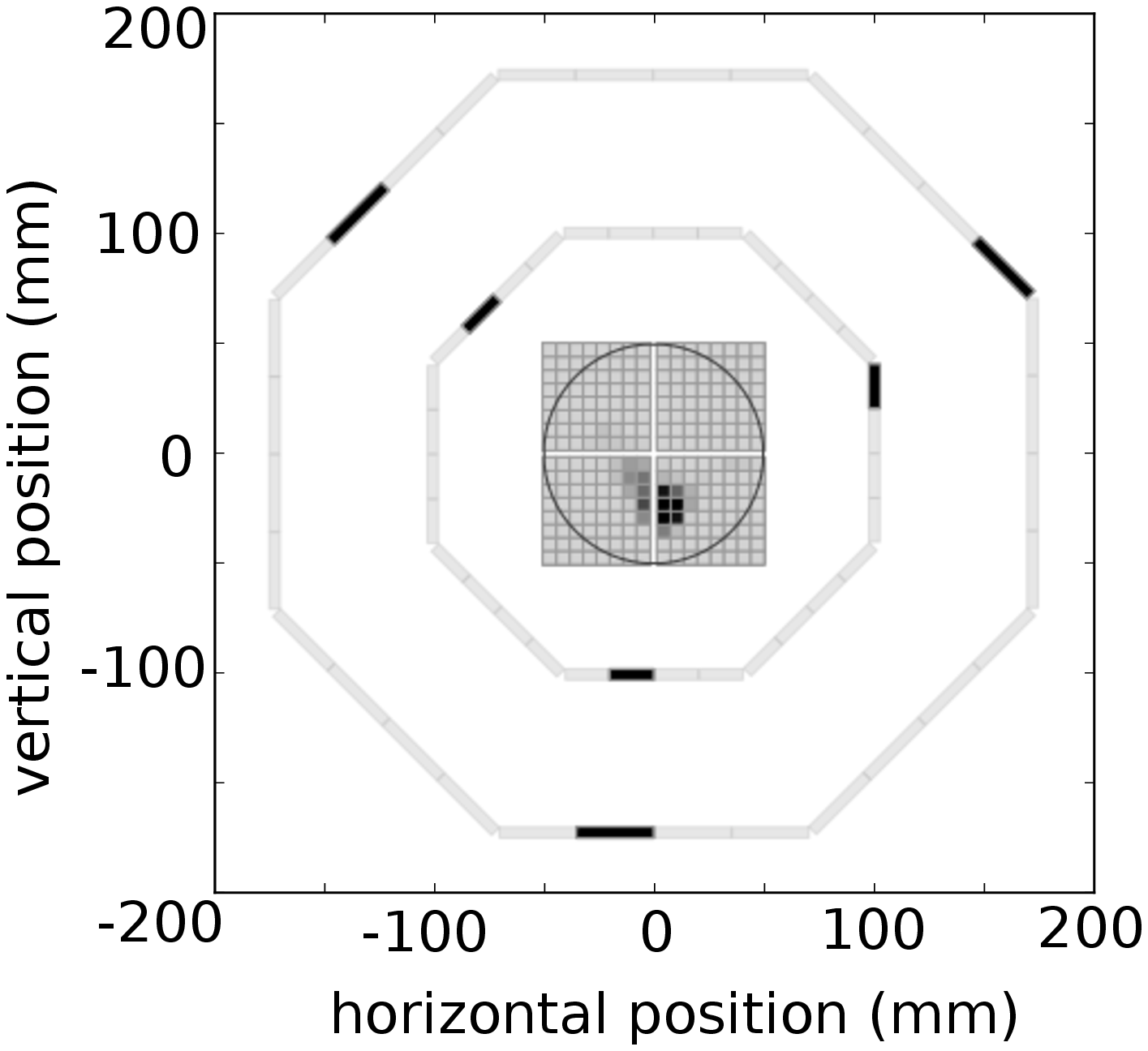} 
    \includegraphics[width=4.8cm]{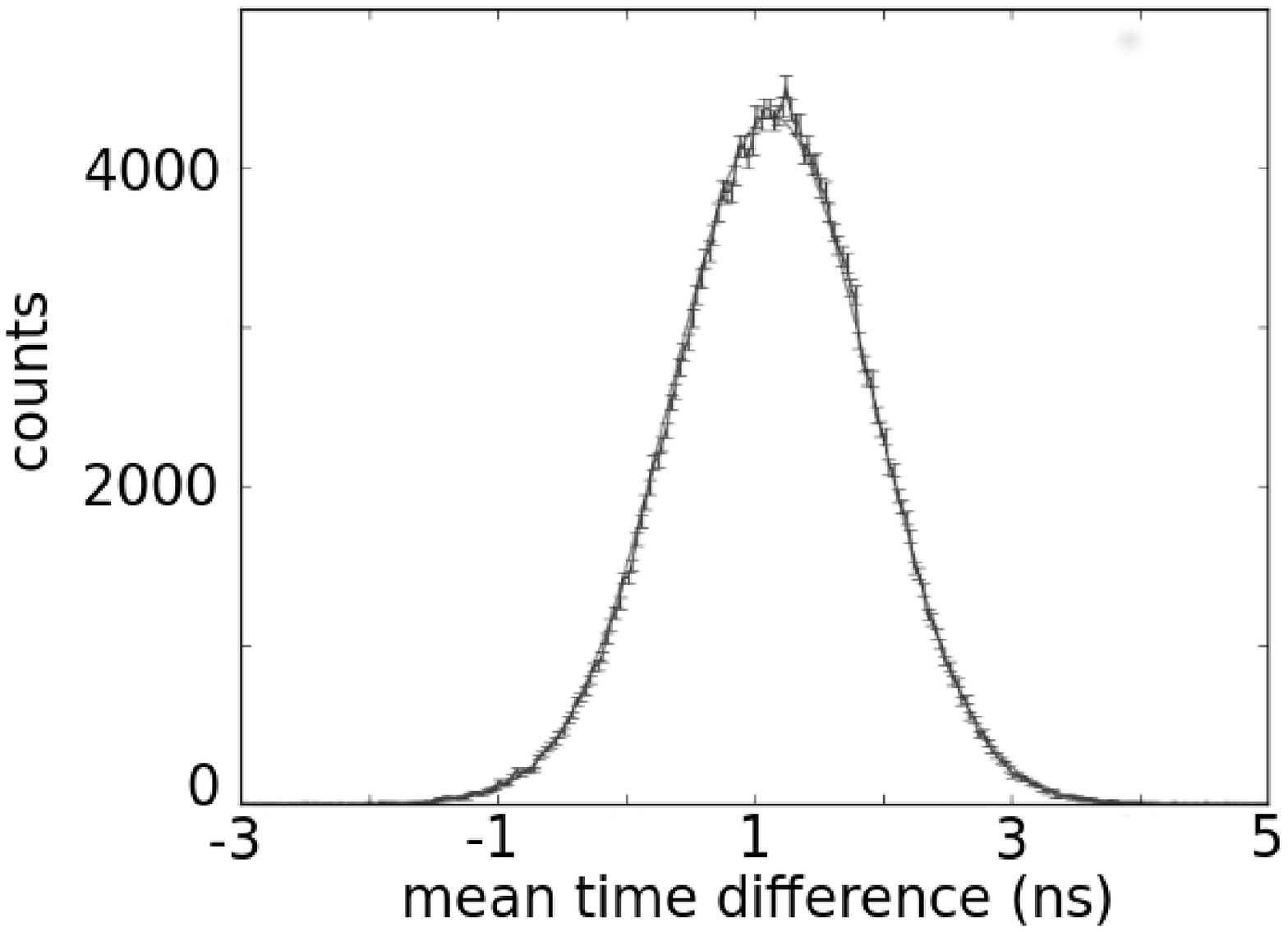}
\end{array}$
\caption{\label{fig:events} Left: cross section of the detector showing a typical cosmic event. Center: cross section of the detector showing an antiproton annihilation event, recorded during beam time in 2015. Right: preliminary histogram of mean time differences of the outer hodoscope for cosmic events.}
\end{figure}

The tracking detector is composed of two layers of segmented hodoscopes, each made of 32 plastic scintillating bars. The hodoscope covers 50$\%$ of the solid angle and its dimensions are shown in Fig.\ \ref{fig:hodor}. Light guides are attachd on both sides of the bars to match the detecting area of the silicon photomultipliers. These are read using self developed front-end electronics,\cite{ifes} and Caen V1742 waveform digitisers. The recorded data is then analyzed using a self developed modular waveform analysis library.\cite{wavelib}
The angular track resolution is defined by the geometry of the bars and via the two sided read-out of the bars. The hit position in beam direction can be determined by combining the information of timing and amplitude of the recorded waveforms. For the outer hodoscope bars the resolution was determined to be 73$\pm$3~mm FWHM and for the inner bars  59$\pm$4~mm FWHM.\cite{detector} This allows rudimentary traking in 3D in order to discriminate between straight tracks created by cosmics and tracks with a kink due to antiproton annihilations.
A relativistic particle with a velocity close to the speed of light needs approximately 1~ns to travel a distance of 30~cm. With a time of flight resolution of $<$ 600 ps it is therefore possible to separate events stemming from inside due to antiproton annihilations from those passing the detector from outside. See Fig.\ \ref{fig:events}. The time of flight resolution was measured with cosmic particles in the laboratory by calculating the mean time difference of two bars resulting in 551$\pm$5~ps FWHM for outer bars and 497$\pm$3~ps for inner bars.\cite{detector} This was remeasured in the complete setup at the AD during beam time this year. See the right plot of Fig.~\ref{fig:events} which displays the preliminary distribution of the mean time difference histogram of the outer hodoscope for cosmic events. The results show a slightly higher FWHM of about 900~ps which is under investigation and can most likely be explained by the longer signal cables in the full setup and differences in the data acquisition system.

\section*{Acknowledgments}
This work was funded by the European Research Council under European Union's Seventh Framework Programme (FP7/2007-2013)/ERC Grant Agreement (291242) and the Austrian Ministry of Science and Research, Austrian Science Fund (FWF): DK PI (W 1252). Hardware development and manufacturing was performed by the SMI workshop.

\end{document}